\documentclass[aps,prd,preprint,floatfix,nofootinbib,showpacs]{revtex4-1}
\pdfoutput=1
\usepackage{graphicx,color,epstopdf,amsmath}
\usepackage{hyperref}

\begin{document}
\title{Interpretations of the ATLAS Diboson Anomaly}
\author{ Kingman Cheung$^{1,2,3}$, Wai-Yee Keung$^{4,1}$, Po-Yan Tseng$^{2}$,
and Tzu-Chiang Yuan$^{5,1}$}
\affiliation{
$^1$ Physics Division, National Center for Theoretical Sciences, Hsinchu,
Taiwan \\
$^2$Department of Physics, National Tsing Hua University, 
Hsinchu 300, Taiwan \\
$^3$Division of Quantum Phases \& Devices, School of Physics, 
Konkuk University, Seoul 143-701, Republic of Korea \\
$^4$Department of Physics, University of Illinois at Chicago, IL 60607 USA \\
$^5$Institute of Physics, Academia Sinica, Nangang, Taipei 11529, Taiwan
}

\renewcommand{\thefootnote}{\arabic{footnote}}
\date{\today}

\begin{abstract}
Recently, the ATLAS Collaboration recorded an interesting anomaly
in diboson production with excesses at the diboson invariant 
mass around 2 TeV in boosted jets of 
all the $WZ$, $W^+ W^-$, and $ZZ$ channels.
We offer a theoretical interpretation 
of the anomaly using a phenomenological right-handed model with extra $W'$ and
$Z'$ bosons. Constraints from narrow total decay widths, 
dijet cross sections, and $W/Z + H$ production
 are taken into account. 
We also comment on a few other possibilities.
\end{abstract}

\maketitle

\section{Introduction}
Recently, the ATLAS Collaboration \cite{atlas} reported an experimental
anomaly in diboson production with apparent excesses 
 in boosted jets 
of the $W^+W^-$, $W^\pm Z$, and $ZZ$ channels at around 2 TeV invariant mass
of the boson pair. 
\footnote
{
The CMS Collaboration also saw a moderate excess around 2 TeV in the
boosted jets of $W^+W^-$, $W^\pm Z$, and $ZZ$ \cite{cms-anom}.
}
It is intriguing because the excesses are all around
2 TeV. The local excesses are at 
$3.4$, $2.6$, and $2.9$ $\sigma$ levels for $WZ$, $W^+ W^-$, and
$ZZ$ channels, respectively
(though the global significance of the discrepancy in the $WZ$
channel is only $2.5\sigma$.)
 The experiment used the method of 
jet substructure to discriminate the hadronic decays of the $W$ and $Z$
bosons from the usual QCD dijets. The advantage is that the hadronic
decays of $W$ and $Z$ afford much larger branching ratios for more
events. However, the jet masses of the $W$ and $Z$ bosons have large overlaps
such that the $Z$ boson may be
misinterpreted as a $W$ boson, and vice versa.
Nevertheless, we attempt to provide a logical explanation for the
anomaly.

The anomaly leads to a logical explanation that there exist some exotic 
particles in some forms of multiplets or under some symmetries (because 
they have similar masses) with relatively narrow widths decaying into 
diboson channels. 
In this note, we propose a phenomenological left-right model that consists
of an extra $SU(2)_R$ gauge group with $(W',Z')$ bosons. 
Initially, we first perform phenomenological studies of the $W'$ and $Z'$ 
boson with respect to the data separately. At the end, we shall 
give a more unified picture of the
$W'$ and $Z'$ bosons coming from a single $SU(2)_R$ group.

The $W'$ 
boson couples to the right-handed fermions with a strength $g_R$, which 
need not be the same as the weak coupling $g$. 
The $W'$ boson can then
be produced via $q\bar q'$ annihilation. Since the $W'$ boson is at 2 TeV,
the production is mainly via valence quarks and so we anticipate
the production cross section of $W'^+$ is roughly two times as large as
the $W'^-$ cross section 
at the LHC. The $W'$ boson can mix with the SM $W$ boson
via a mixing angle, say, $\sin\phi_w$ so that the $W'$ boson can decay 
into $WZ$ with a mixing-angle suppression and right-handed fermions. 
We shall show that the $W'$ decay into $WZ$ dominates if the mixing angle is
larger than $10^{-2}$.
Therefore, it can explain the excess in the $WZ$ channel
without violating the 
leptonic cross sections \cite{atlas-wp,cms-wp,atlas-zp,cms-zp} 
and the dijet-mass search at the LHC \cite{atlas-dj,cms-dj}.

The discussion of the $Z'$ boson follows closely that of the $W'$ boson.
It is produced via $q \bar q$ annihilation with a coupling strength $g_R$.
The $Z'$ boson mixes with the SM $Z$ boson via another mixing angle
$\phi_z$, and then decays into $W^+ W^-$ to explain the excess in
the $W^+ W^-$ channel.
We adopt a simplified form that the $Z'$ only couples to the
right-handed fermions, though in general it couples to both
left- and right-handed fermions.

There are, in general, a few important constraints that restrict the form 
the $W'$ and $Z'$ models: 
(i) electroweak (EW) precision measurements, (ii) leptonic
decays of $W'$ and $Z'$, and (iii) dijet production cross sections,
plus $WH$ and $ZH$ production that are specific to the current work.
The EW precision constraints mainly come from the deviations in the
properties of the observed $W$ and $Z$ bosons through the mixings
between $W$ and $W'$ bosons, and between $Z$ and $Z'$ bosons.
The measured properties of the $W$ boson restrict the mixing
angle between the $W$ and $W'$ boson to be 
$\phi_w \alt 1.3 \times 10^{-2}$ \cite{pdg}, 
which is the approximate size of the
mixing that is required to explain the diboson anomaly. 
On the other hand, the constraint on the mixing angle between the $Z$ and 
$Z'$ bosons is much stronger.  The updated limits for various $Z'$ 
models, in which the $Z'$ boson has direct couplings to SM particles,
are of order $10^{-3}$.  This is somewhat smaller than the values required 
to explain the diboson anomaly.  
One possibility to relax this constraint is to
employ the leptophobic $Z'$ model, which is achievable in a number of 
GUT models \cite{paul}.
In such a case, the constraint on the mixing angle can be relaxed to
$8 \times 10^{-3}$ \cite{paul}, which is close to 
the value required to explain the diboson anomaly.
Therefore, we shall employ the leptophobic $Z'$ model in this work.
Furthermore, in the leptonic decays of $W'^+ \to e^+ \nu_R$ we assume
the right-handed neutrinos are heavy enough that the leptonic decays
of the $W'$ boson are also closed.

Note that we take the excess in the $ZZ$ channel as either a fluctuation
or the mis-interpretation because of the overlap between the $W$ and $Z$ 
dijets. On the theory side, it is very difficult to have
a spin-1 particle to decay significantly
into $ZZ$, e.g., the $Z'$ boson \cite{Keung:2008ve,Modak:2014zca} or a techni-rho
meson. It is possible to have a spin-0 Higgs-like boson to decay into 
$ZZ$. However, we found that the production cross section for a 2 TeV
Higgs-like boson via gluon fusion is too small to explain the
excess in the $ZZ$ channel. Therefore, we take the liberty to ignore
the excess in the $ZZ$ channel.

The organization 
of this note is as follows. In the next section, we describe the 
interactions of the $W'$ and $Z'$ bosons, and mixing with the SM 
$W$ and $Z$ bosons. In Sec. III, we calculate 
the dijet cross sections to compare with the most updated limits 
from ATLAS \cite{atlas-dj} and CMS \cite{cms-dj}. In Sec. IV, we calculate
the cross sections
of $pp \to W' \to WZ$ and $pp \to Z' \to W^+ W^-$ and compare to
the ATLAS data. 
In Sec. V, we give a more unified picture that the
$W'$ and $Z'$ bosons come from a single $SU(2)_R$.
We conclude in Sec. VI.

At the last stage of this work, the authors came across Ref.~\cite{hisano}
with a similar idea, and Refs.~\cite{tech1,tech2} in the framework
of strong dynamics.
There are some existing constraints in literature for models with extra
$SU(2)$ \cite{new}, especially the dilepton
constraint from the LHC experiments. We shall consider the
dilepton constraint, as well as the dijet constraint using the most
recent data from the LHC.

\section{Interactions of the $W'$ and $Z'$ bosons and their decays}

\subsection{The $W'$ boson}
The extra $W_2$ boson arises from the right-hand $SU(2)_R$. 
The right-handed fermions are arranged in isospin doublet, e.g,
$(u_R, d_R)^T,\; (\nu_R, e_R)^T$, where $\nu_R$ is the right-handed 
neutrino. 
The interactions of the $W_2$ with fermions are given by
\begin{eqnarray}
{\cal L} &\supset& -\frac{g_R}{\sqrt{2}} \bar f' \gamma_\mu P_R f \, W_2^\mu 
\label{eq1}
\end{eqnarray}
where $P_{L,R}  = (1 \mp \gamma^5)/2$ and $g_R$ is the coupling strength,
which need not be the same as the left-handed coupling $g$ but
should be of a similar size.  The $W_1$ and $W_2$ denote
the interaction eigenstates, which rotate into the mass eigenstates
$W$ and $W'$ via a mixing angle $\phi_w$ ($W$ then represents the 
observed $W$ boson at $80.4$ GeV and the $W'$ is the hypothetical 2 TeV
boson):
\begin{equation}
 \left (
  \begin{array}{c}
    W_1 \\
    W_2   \end{array} \right )=  \left(\begin{array}{lr}
               \cos\phi_w & -\sin\phi_w \\
               \sin\phi_w & \cos\phi_w  \end{array} \right )\;
 \left(  \begin{array}{c}
    W \\
    W'   \end{array} \right ) \;.
\end{equation}
Current EW constraints on the $W$-$W'$ mixing angle mainly 
come from modifying the properties of the observed $W$ boson. 
The measurements put a limit about $1.3\times 10^{-2}$ \cite{pdg}
on the mixing angle $\phi_w$,  which is more or less consistent with
the values that we use in this study (see Fig.~\ref{fig-para-w}).
We shall show that such a small mixing angle of order $O(10^{-2})$
is enough to explain the narrow width of the $W'$ bump and the excess in
the $WZ$ production cross section.

On the other hand, due to the mixing with the SM $W$ boson, the heavy 
$W'$ couples to $WZ$ with a coupling strength 
$(g\cos\theta_w ) \sin\phi_w$,
where $(g\cos\theta_w)$ is the usual coupling constant in the $WWZ$ vertex.
On the other hand, the $W'$ couples to $WH$ with 
a full tree-level strength $(g M_W )\sin\phi_w
\cos^2\theta_w  \frac{M_{W'}^2}{M_W^2}$.
\footnote
{
The mixing angle $\phi_w$ between the $W$ and $W'$ originates from the 
off-diagonal mass matrix entry, which also gives the tree-level unsuppressed
coupling for $W'$-$W$-$H$. We found that $\phi_w$ scales as $M_W^2/M_{W'}^2$
and the coupling for $W'WH$ scales as $g M_W \phi_w M_{W'}^2/M_{W}^2$.
}
Now we can write down the relevant vertices of the $W'$ used in this 
work ($\cos\phi_w \simeq 1$),
\begin{eqnarray}
{\cal V}_{W'ff'}  & : & - \frac{g_R}{\sqrt{2}} \bar f' \gamma^\mu P_R f \, 
   \epsilon_\mu (p_{W^{'+}}) \;\;, 
 \nonumber \\
{\cal V}_{W'WZ}   & : & + g \cos\theta_w \sin\phi_w \left[
  (p_{W^{'+}} - p_{W^-})^\beta g^{\mu\alpha} + (p_{W^-} - p_Z)^\mu g^{\alpha\beta} 
 + (p_Z - p_{W^{'+}})^\alpha g^{\mu\beta} \right ] \, \nonumber \\
  && \times  \epsilon_\mu(p_{W^{'+}}) \;
   \epsilon_\alpha(p_{W^-}) \, \epsilon_\beta (p_Z) \;\; , \nonumber \\
{\cal V}_{W'WH}  & : & + g M_W \sin\phi_w \, 
  \left( \cos^2\theta_w  \frac{M_{W'}^2}{M_W^2} \right ) \, 
g^{\mu\alpha}\, \epsilon_\mu(p_{W^{'+}}) \, 
       \epsilon_\alpha( p_{W^-}) 
\; \; , \label{lag}
\end{eqnarray}
where $p_{W^{'+},W^-,Z}$ denote the 4-momenta of the $W^{'+},W^-,Z$ bosons 
going into the vertex and
$\epsilon(p_{W^{'+}})$,  $\epsilon(p_{W^-})$, and $\epsilon(p_{Z})$ denote
the corresponding polarization 4-vectors.

\begin{figure}[t!]
\includegraphics[width=3in]{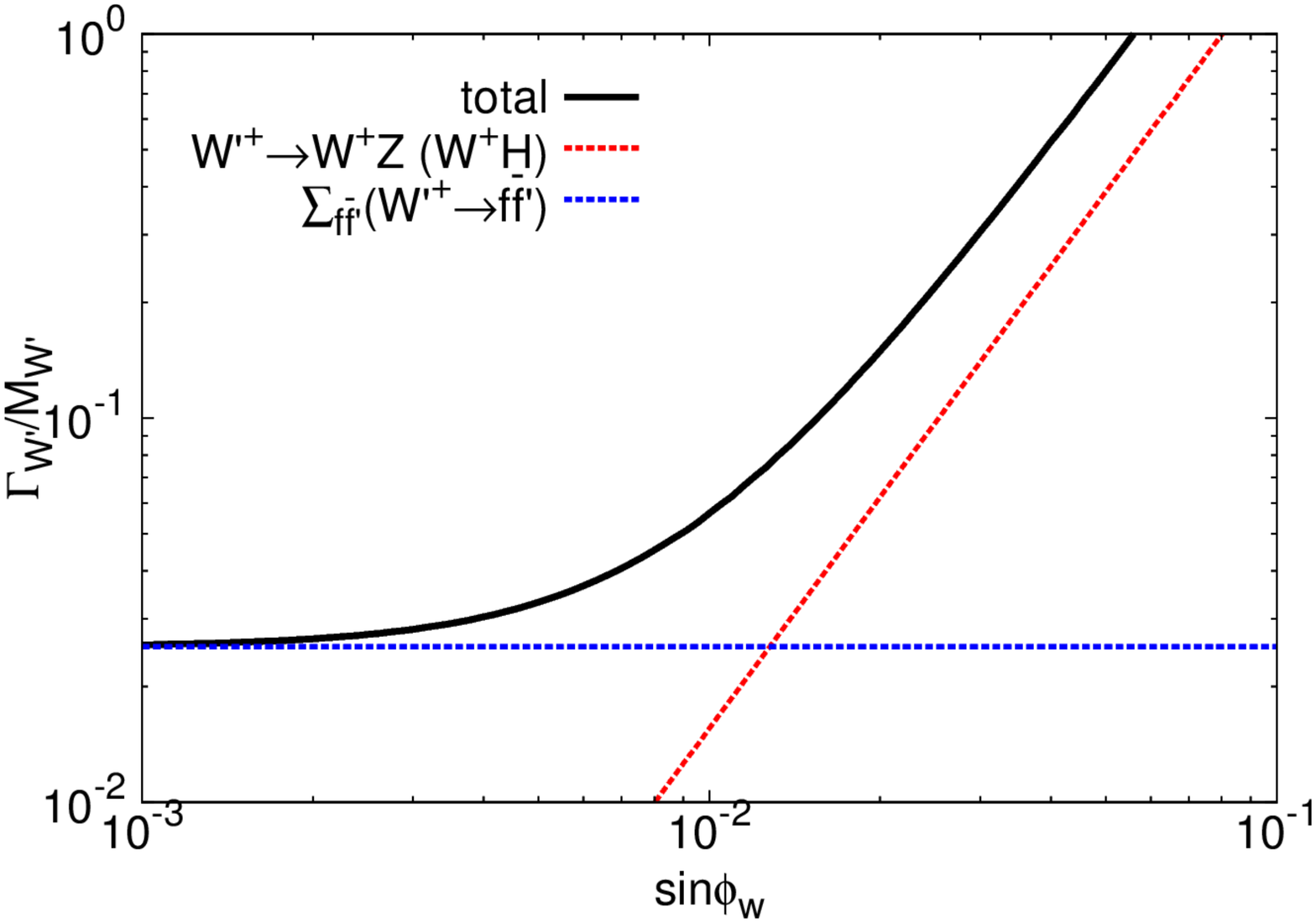}
\includegraphics[width=3in]{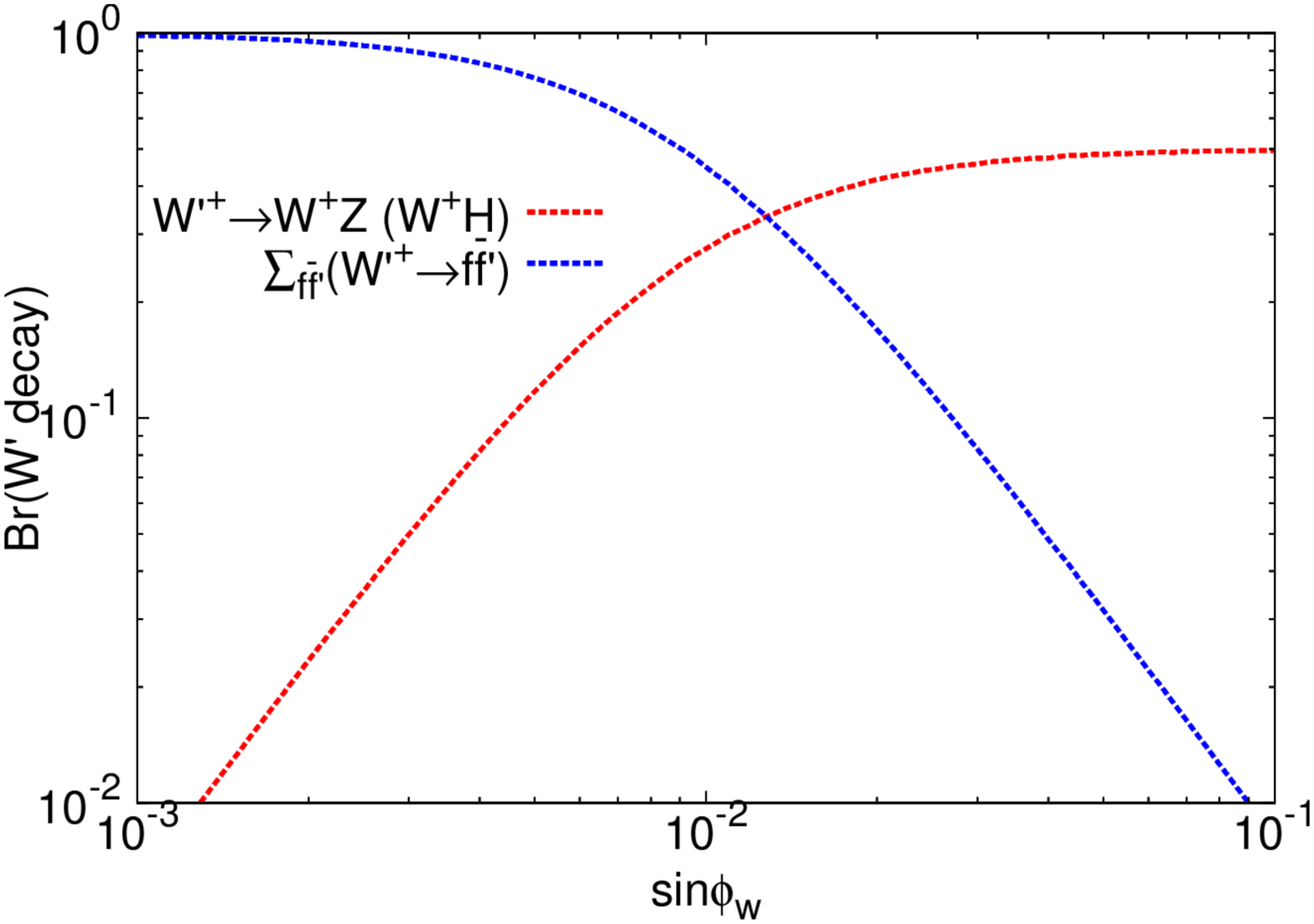}
\caption{\small \label{fig1}
(Left) Total decay width and partial widths  of the 2 TeV $W'$ boson
 versus the sine of the mixing angle $\phi_w$. 
(Right) The corresponding branching ratios.
Note that the $W^+H$ appears to be the same as $W^+ Z$ in the figure.
Here we take $g_R = g$.}
\end{figure}

The partial decay width for $W'\rightarrow f\bar{f'}$ is given by,
in massless limit of $f,f'$
\begin{equation}
\Gamma_{W'\rightarrow f\bar{f'}} = 
  \frac{N_f g_R^2 M_{W'} }{48\pi} \;,
\end{equation}
where $N_f = 3\;(1)$ for quarks (leptons).
Here we also assume that the  
right-handed neutrinos are so heavy that $W'^+ \to e^+ \nu_R$ 
is kinematically not allowed.  Therefore, the leptonic decay modes of
$W'$ are closed.
The partial width into $WZ$ is \cite{altarelli}
\begin{eqnarray}
\Gamma_{W'^+ \rightarrow W^+Z} &=& {\rm{sin}}^2\phi_w 
\left( \frac{g^2 \cos^2\theta_w}{192\pi} \frac{M^5_{W'}}{M^2_W M^2_Z}  
\right) \;
\left[ \left(1-\frac{M^2_Z}{M_{W'}^2} - \frac{M^2_W}{M^2_{W'}} \right)^2-
  \frac{4M^2_W M_Z^2 }{M^4_{W'}}
 \right]^{3/2} \nonumber \\
&& \times \left[1+10\left( \frac{M^2_Z+M^2_W}{M^2_{W'}} \right)+
 \frac{M^4_Z+M^4_W+10M^2_ZM^2_W}{M^4_{W'}} \right]\,.
\end{eqnarray}
It is easy to see that in the $W'WZ$ vertex in Eq.~(\ref{lag}) the 
momentum-dependent parts will get enhancement at high energy.
Another decay channel of $W'$ is $W' \to W H$, the partial width of 
which is given by
\begin{eqnarray}
\Gamma_{W'^+ \to W^+ H} &=& \sin^2 \phi_w \left(
\frac{g^2}{192 \pi} \cos^4\theta_w \frac{M_{W'}^5}{M_W^4}
   \right) \, \left[
\left ( 1+ \frac{M_W^2} {M_{W'}^2} - \frac{M_H^2}{M_{W'}^2  } \right )^2 
  + 8 \frac{M_{W}^2}{M_{W'}^2} \right]  \nonumber \\
&\times & \, 
\left[ \left( 1- \frac{M_W^2}{M_{W'}^2} - \frac{M_H^2}{M_{W'}^2  } \right )^2 
  -4 \frac{M_{W}^2 M_H^2 }{ M_{W'}^4 } \right ]^{1/2} \; .
\end{eqnarray}
Note that 
due to the Equivalence Theorem, $\Gamma(W'^+ \to W^+ H) \simeq \Gamma(
W'^+ \to W^+ Z)$ to the leading order in $1/M^2_{W'}$.

In order to avoid a too-broad resonance structure for the $W'$ boson we require
\begin{equation}
  \Gamma_{\rm tot} (W'^+)= \Gamma_{W'^+ \to W^+ Z} + 
\Gamma_{W'^+ \to W^+ H} + 
\sum_{f\bar f' = u\bar d, c\bar s, t\bar b} \Gamma_{W'^+ \to f\bar f'}  \label{br-wp}
\end{equation}
to be less than one-tenth of the $W'$ mass.
We show in Fig.~\ref{fig1} the total width of the $W'$ boson versus
the sine of the mixing angle, and the corresponding branching ratios.
Note that the $W^+H$ appears to be the same as $W^+ Z$ in the figure.
It is clear that the total decay width
grows with $\sin\phi_w$ rapidly. Therefore, the requirement of $\Gamma_{\rm tot}
(W') \alt M_{W'}/10$ gives
\begin{equation}
\sin\phi_w \alt 1.5\times 10^{-2} \;.
\end{equation}
We show in Table~\ref{tab1} a few representative values of $\sin\phi_w$
for the partial widths into $WZ$ or $WH$,
and $\sum f\bar f'$ and the total width 
of the $W'$ boson. Here we assume $g_R = g$ for simplicity. Note that
the total width has
only very mild dependence on $g_R$. 
%Even though in the
%next section, due to the tight constraints on leptonic decays of 
%the $W'$ and $Z'$ bosons we have to choose $g_R = g/2$ or smaller,
%the requirement on the mixing angle $\sin\phi_w$ remains almost the 
%same.

\begin{table}[thb!]
\caption{\small \label{tab1}
The partial widths into $WZ/WH$ and $\sum f\bar f'$ and the total width 
of the $W'$ boson for a few representative values of $\sin\phi_w$.
We assume $\Gamma_{W'^+ \to W^+ Z} = \Gamma_{W'^+ \to W^+ H}$
and set $g_R = g$ for simplicity. }
\begin{ruledtabular}
\begin{tabular}{ccccc}
Case & $\sin\phi_w$ & $\Gamma_{W'^+ \rightarrow W^+Z/{W^+ H}}$ (GeV) & 
$\sum_{f\bar{f'}} \Gamma_{W'^{+}\rightarrow f\bar{f'}}$  (GeV)
& $\Gamma_{W'^{+}}$  \\
\hline
1 & $8.901\times 10^{-3}$ & 24.63 & 50.74 & $M_{W'}/20$  \\
\hline
2 & $1.549\times 10^{-2}$ & 74.63 & 50.74 & $M_{W'}/10$  \\
\hline
3 & $2.370\times 10^{-2}$ & 174.6 & 50.74 & $M_{W'}/5$ \\
\hline
4 & $3.148\times 10^{-2}$ & 308.0 & 50.74 & $M_{W'}/3$  \\
\hline
5 & $3.908\times 10^{-2}$ & 474.6 & 50.74 & $M_{W'}/2$ 
\end{tabular}
\end{ruledtabular}
\end{table}

\subsection{The $Z'$ boson}

We repeat the exercise for the $Z'$ boson. The interactions of the $Z_2$ with
the SM fermions are given by
\begin{equation}
\label{z-f}
{\cal L} \supset - \bar f \gamma_\mu (g_{f,r} P_R + g_{f,l} P_L) f 
\, Z_2^\mu \;\;.
\end{equation}
The SM $Z_1$ boson mixes with $Z_2$ via a mixing angle $\phi_z$ into
the mass eigenstates $Z$ and $Z'$:
\begin{equation}
 \left (
  \begin{array}{c}
    Z_1 \\
    Z_2   \end{array} \right )=  \left(\begin{array}{lr}
               \cos\phi_z & -\sin\phi_z \\
               \sin\phi_z & \cos\phi_z  \end{array} \right )\;
 \left(  \begin{array}{c}
    Z \\
    Z'   \end{array} \right ) \;.
\end{equation}

Unlike the $W$-$W'$ mixing the EW constraints for the $Z$-$Z'$ 
mixing angle are much stronger, because of all the
precision measurements at LEP.  The updated limit for various $Z'$ 
models, in which the $Z'$ boson has direct couplings to SM particles,
is of order $10^{-3}$ \cite{paul}. This is somewhat smaller than the values 
required to explain the diboson anomaly.
One possibility to relax this constraint is to
employ the leptophobic $Z'$ model, which is achievable in GUT models.
In such a case, the constraint on the mixing angle can be relaxed to
$8 \times 10^{-3}$ \cite{paul}, which is not far from the values required to
solve the diboson anomaly. We shall therefore assume leptophobic couplings
of the $Z'$ boson.
When the mixing angle is of that small size, the $Z'$ boson has 
a narrow width.

\begin{figure}[th!]
\includegraphics[width=4in]{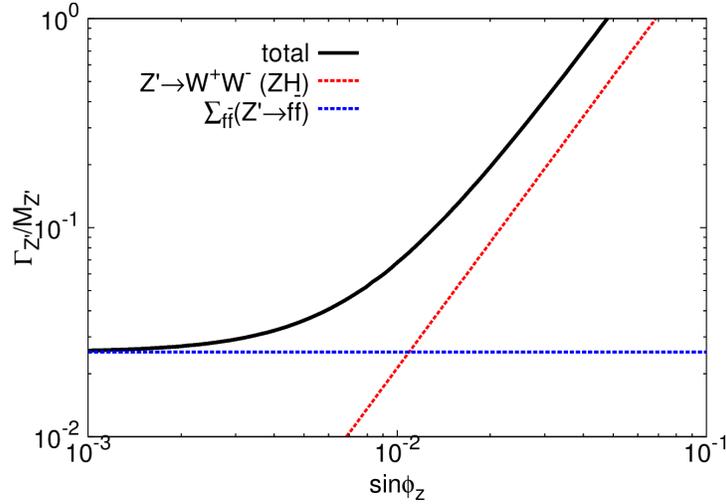}
\caption{\small \label{fig-z}
Total decay width and partial widths  of the 2 TeV $Z'$ boson
versus the sine of the mixing angle $\phi_z$. 
Note that the $ZH$ appears to be the same as $W^+ W^-$  in the figure.
Here we take $g_R = g$ for simplicity.}
\end{figure}

The $Z'$ boson then couples with a strength proportional
to $g_{f,r/l}$
to the SM quarks, but at a strength suppressed by the 
mixing angle $\sin\phi_z$ to the $W^+ W^-$.
However, the $Z'$ boson couples to $ZH$ with a full tree-level strength
for a reason similar to the $W'$ boson.
Now we can write down the relevant vertices of the $Z'$ 
used in this work taking $\cos \phi_z \simeq 1$:
\begin{eqnarray}
{\cal V}_{Z'ff} & : &  - \bar f  \gamma^\mu 
       (g_{f,r} P_R + g_{f,l} P_L) f \, 
   \epsilon_\mu (p_{Z'}) \;\; ,
 \nonumber \\
{\cal V}_{Z'WW} & : &  +g \cos\theta_w \sin\phi_z  \left[
  (p_{Z'} - p_{W^+} )^\beta g^{\mu\alpha} + (p_{W^+} - p_{W^-})^\mu g^{\alpha\beta} 
 + (p_{W^-} - p_{Z'})^\alpha g^{\mu\beta} \right ] \, \nonumber \\
  && \times  \epsilon_\mu(p_{Z'}) \;
   \epsilon_\alpha(p_{W^+}) \, \epsilon_\beta (p_{W^-} ) \;\;, \nonumber \\
{\cal V}_{Z'ZH} & : & +\frac{g}{ \cos\theta_w} M_Z \sin\phi_z 
 \left( \frac{M_{Z'}^2 }{M_Z^2 } \right ) \,
g^{\mu\alpha}\, \epsilon_\mu(p_{Z'})
 \, 
       \epsilon_\alpha( p_Z) 
\; \; . \label{lag_z}
\end{eqnarray}
The partial widths into $f\bar f$, $W^+ W^-$, and $ZH$ are given by
\begin{eqnarray}
\Gamma_{Z' \to f \bar f} &=& N_f 
\frac{ g_{f,r}^2+g_{f,l}^2}{ 24 \pi} \, M_{Z'} \; , \nonumber  \\
\Gamma_{Z' \to  W^+ W^- } &=& \sin^2\phi_z 
\left( \frac{g^2 \cos^2\theta_w }{192\pi } \frac{M^5_{Z'}}{M^4_W}  \right ) \,
\left( 1- \frac{4M_W^2}{M_{Z'}^2 }\right )^{3/2}\,
\left( 1 + 20 \frac{M_W^2}{M_{Z'}^2 } +12 \frac{M_W^4}{M_{Z'}^4 } \right)\; ,
\nonumber \\
\Gamma_{Z' \to  ZH } &=& 
\sin^2 \phi_z \left(
\frac{g^2 \cos^2\theta_w }{192 \pi } \frac{M_{Z'}^5}{M_W^4} 
   \right) \, \left[
\left ( 1+ \frac{M_Z^2} {M_{Z'}^2} - \frac{M_H^2}{M_{Z'}^2  } \right )^2 
  + 8 \frac{M_{Z}^2}{M_{Z'}^2} \right]  \nonumber \\
&\times & \, 
\left[ \left( 1- \frac{M_Z^2}{M_{Z'}^2} - \frac{M_H^2}{M_{Z'}^2  } \right )^2 
  -4 \frac{M_{Z}^2 M_H^2 }{ M_{Z'}^4 } \right ]^{1/2} \label{br-zp} \; .
\end{eqnarray}
In the high energy limit, $\Gamma (Z' \to W^+ W^-) \simeq
\Gamma(Z' \to ZH)$.
The total decay width of the $Z'$ boson is obtained by summing all the 
above partial widths. We show the total decay width and partial widths
of the $Z'$ boson in Fig.~\ref{fig-z}.  The requirement for $\Gamma_{Z'} /
M_{Z'} < 0.1$ implies 
 $\sin\phi_z \alt 1.5 \times 10^{-2}$. 

Note that the couplings of $Z_2$ to quarks
are model dependent (leptonic couplings are zero).
We use the simplified form 
$g_{f,l}=0, g_{f,r}=g_R T^{(2)}_{3,f}$
for the interactions
according to the right-handed current $T^{(2)}_{3}$  as in Eq.~(\ref{z-f}).
Our analysis can be generalized to any specific model  
by scaling the corresponding couplings.

The mixing between the $Z$ and $Z'$ bosons can be generated through 
a Higgs boson charged under both symmetries. The mixing is given by
$\phi_z = C (g_R/g) (M_V/M_{V'})^2$ \cite{paul-luo}, 
which $C$ can be a definite number or 
spanned over a range depending on the ratios of the Higgs VEVs. Given that
$ (M_V / M_{V'}) \sim 10^{-2}$ and $(g_R / g) \sim 0.3 - 1$, the 
mixing angle $\phi_z \sim 10^{-3} - 10^{-2}$. The mixing angle that we find
in this work is about $2\times 10^{-3} - 10^{-2}$ and is 
mostly consistent with the natural value.

\section{Limits from dijet production and others}

\begin{figure}[t!]
\includegraphics[width=4in]{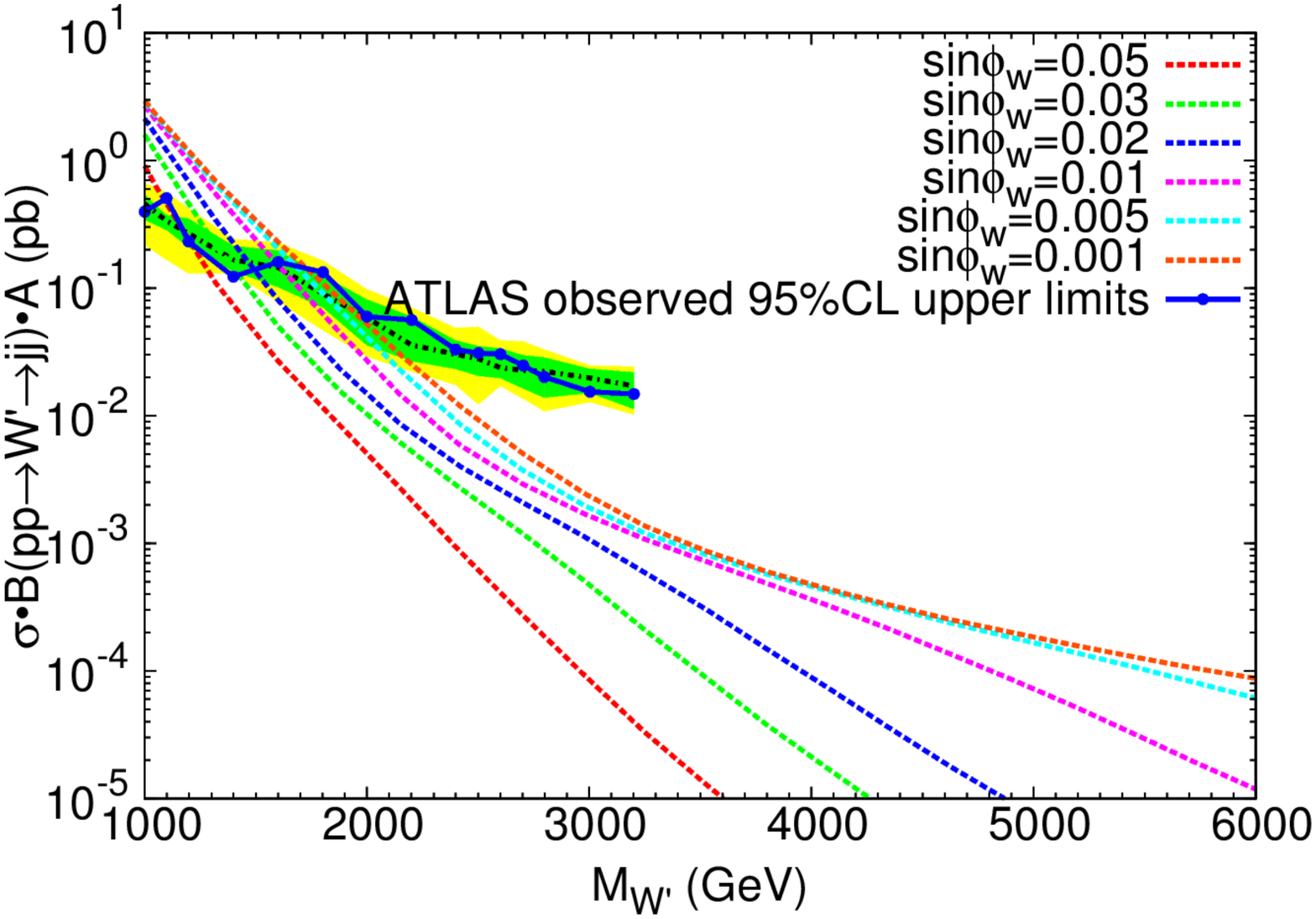}
\includegraphics[width=4in]{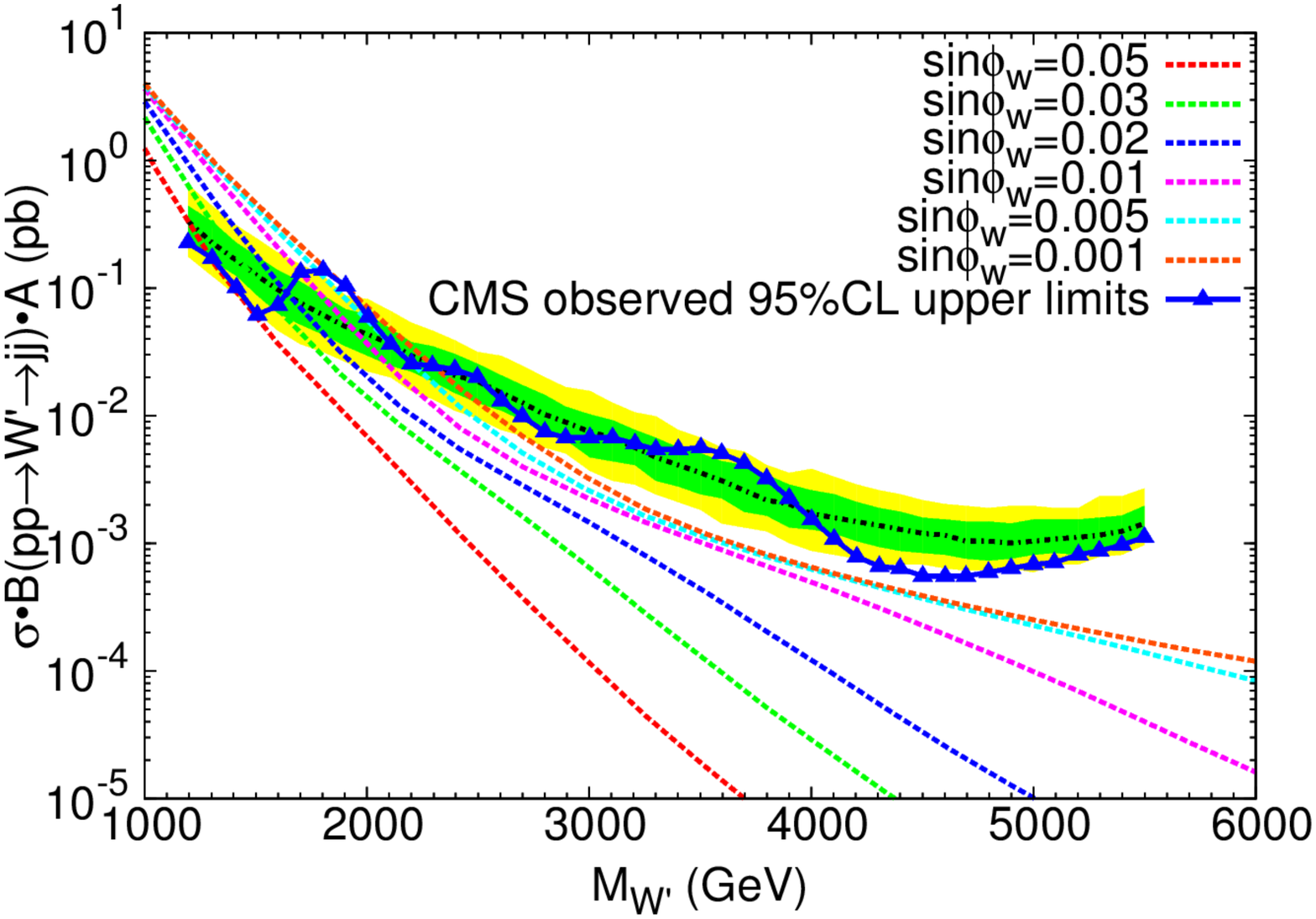}
\caption{\small \label{dijet}
Dijet production cross sections $\sigma\cdot B (pp\to W'^\pm \to jj) 
\cdot A$ versus the mass of the $W'$ boson for
a few values of $\sin\phi_w$, where $A$ is the acceptance from the experiments.
Here we take $g_R = 0.6$. The ATLAS and CMS 95\% CL upper limits are also 
shown.
}
\end{figure}

\begin{figure}[th!]
\includegraphics[width=4in]{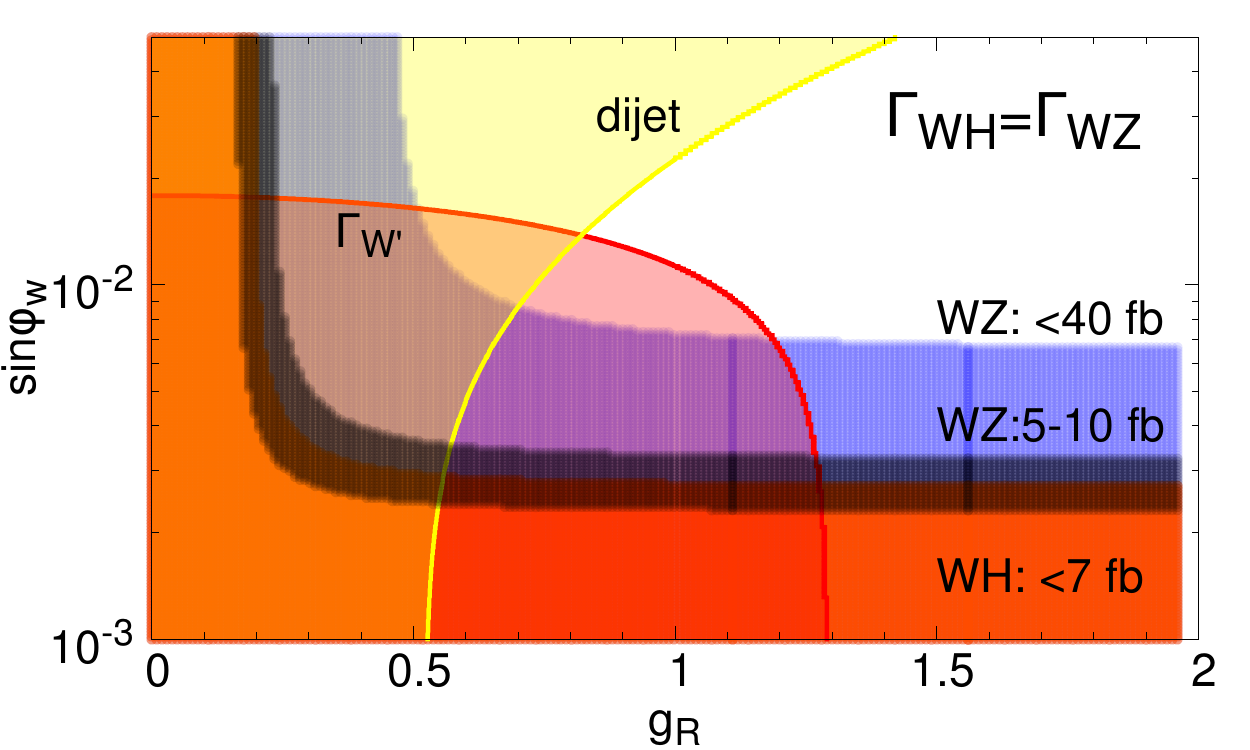}
\caption{\small \label{fig-para-w}
The allowed parameter space in $g_R$ versus $\sin\phi_w$ for the $W'$ 
boson under the constraints:
$\Gamma_{W'} / M_{W'}  < 0.1$,
$\sigma(W') \times B(W' \to jj ) \cdot A < 60 $ fb,
$\sigma(W') \times B(W' \to WZ ) < 40 $ fb, and
$\sigma(W') \times B(W' \to WH ) < 7 $ fb.
}
\end{figure}

Since we have assumed the leptophobic $Z'$ model and that the
right-handed neutrinos are too heavy for $W'^+ \to e^+ \nu_R$ to occur,
the constraints from leptonic cross sections 
\cite{atlas-wp,cms-wp,atlas-zp,cms-zp} can be ignored.
In the following, we first consider the constraints coming
from dijet production via $\sigma(W') \times B(W'\to jj)$ (and 
similarly for $Z'$).
Both the ATLAS \cite{atlas-dj} and CMS \cite{cms-dj} have searched for
resonances decaying into dijets. They pose limits on the
current phenomenological $W'$ and $Z'$ model. We calculate 
$pp \to W'^\pm \to jj$ including the width effect and show the 
production cross sections in Fig.~\ref{dijet},
in which we choose $g_R = 0.6$.
The acceptance factor $A$
for each experiment is read off from the report of ATLAS and CMS.
It is easy to see from both panels that when $\sin\phi_w \agt 5 \times 10^{-3}$
the dijet production cross section at $M_{W'} = 2$ TeV is safe from the
experimental limits.
As $g_R$ further increases, the lower limit on 
$\sin\phi_w$ increases, as shown in Fig.~\ref{fig-para-w}.
Note that for $g_R \alt 0.5$ there is no lower limit on $\sin\phi_w$.
The $Z'$ production cross sections are roughly one half of the $W'$ for the
same mass of 2 TeV. We do not expect $\sigma(Z')\times B(Z'\to jj)$ will 
pose any problems as long as $\phi_z \agt 5\times 10^{-3}$ for $g_R=0.6$.

Yet, there is another constraint mentioned in Ref.~\cite{hisano}:
$\sigma (W') \times B(W' \to WH) < 7$ fb.
Both ATLAS \cite{atlas-wh} and CMS \cite{cms-wh} searched for 
a resonance that decays into a $W/Z$ boson and Higgs boson. The 
95\% CL on $\sigma(W'/Z') \times B(W'/Z' \to W/Z + H) \approx 5-10$ fb.
As shall be seen next the required cross section
for $\sigma (W') \times B(W' \to WZ)$ 
to explain the excess is about $6-7$ fb,
and a similar one for $\sigma (Z') \times B(Z' \to W^+ W^-)$. 
It is therefore safe from the $WH$ and $ZH$ constraints.
Finally, there was another constraint on $Z'$ coming from 
a recent search on $Z' \to W^+ W^-$ via the
semileptonic channel of the $W^+ W^-$ decay and put an upper limit on
$\sigma(Z') \times B(Z' \to W^+ W^-) < 3$ fb at 95\% CL \cite{cms-ww}.

\begin{figure}[th!]
\includegraphics[width=4in]{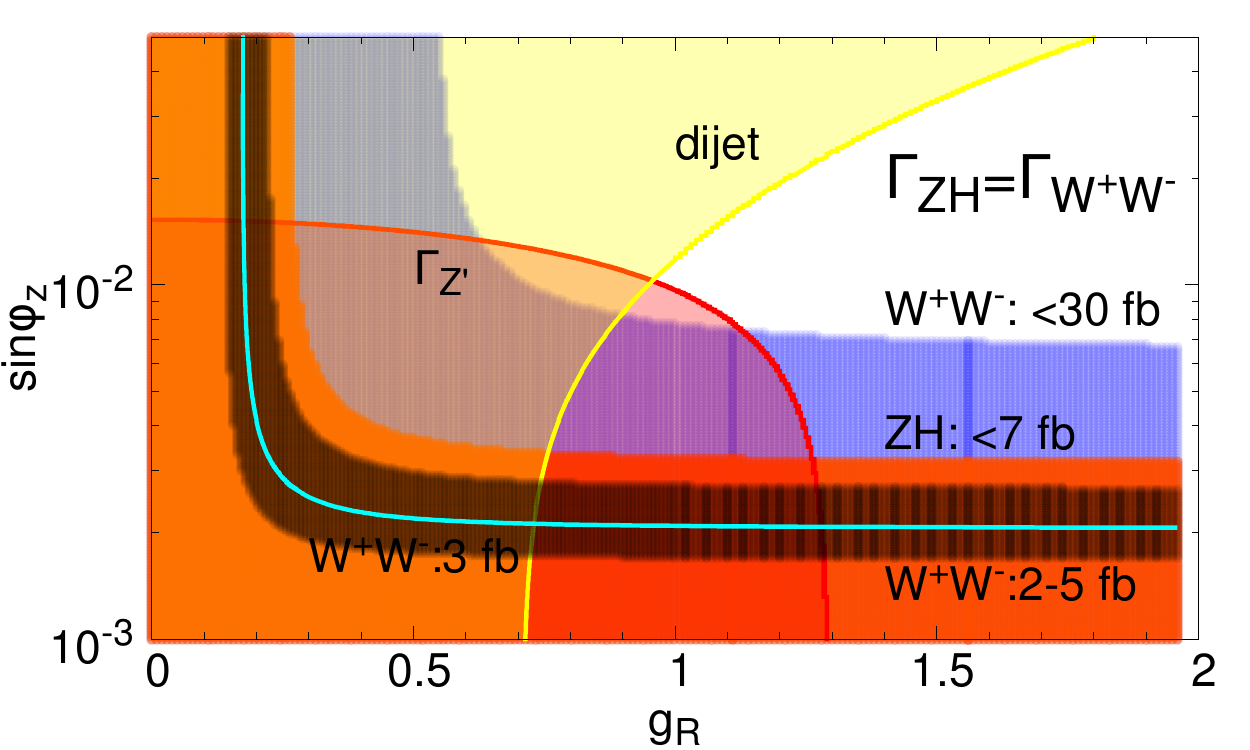}
\caption{\small \label{fig-para-z}
The allowed parameter space in $g_R$ versus $\sin\phi_z$ for the $Z'$ 
boson under the constraints:
$\Gamma_{Z'} / M_{Z'}  < 0.1$,
$\sigma(Z') \times B(Z' \to jj ) \cdot A < 60 $ fb,  
$\sigma(Z') \times B(Z' \to W^+ W^- ) < 30 $ fb, 
$\sigma(Z') \times B(Z' \to ZH ) < 7 $ fb, and
$\sigma(Z') \times B(Z' \to W^+ W^- ) < 3 $ fb.
}
\end{figure}

We summarize in Fig.~\ref{fig-para-w} the allowed parameter space
of $g_R$ versus $\sin\phi_w$ for the 2 TeV $W'$ boson under the 
following constraints:
\begin{enumerate}
\item $\Gamma_{W'} / M_{W'}  < 0.1$,
\item $\sigma(W') \times B(W' \to jj ) \cdot A < 60 $ 
fb \cite{atlas-dj,cms-dj},
\item $\sigma(W') \times B(W' \to WZ ) < 40 $ fb \cite{atlas}, and 
\item $\sigma(W') \times B(W' \to WH ) < 7 $ fb \cite{atlas-wh,cms-wh}.
\end{enumerate}
Similarly, the allowed parameter space in $g_R$ versus $\sin\phi_z$
for the $Z'$ boson with the following constraints is shown in 
Fig.~\ref{fig-para-z}.
\begin{enumerate}
\item $\Gamma_{Z'} / M_{Z'}  < 0.1$,
\item $\sigma(Z') \times B(Z' \to jj ) \cdot A < 60 $ 
fb \cite{atlas-dj,cms-dj},
\item $\sigma(Z') \times B(Z' \to W^+ W^- ) < 30 $ fb \cite{atlas},
\item $\sigma(Z') \times B(Z' \to ZH ) < 7 $ fb \cite{atlas-wh,cms-wh}.
\item $\sigma(Z') \times B(Z' \to W^+ W^- ) < 3 $ fb \cite{cms-ww}.
\end{enumerate}
As seen in both figures the dijet cross section rules out large values 
of $g_R$ while the narrow width requires $\sin\phi_w \alt 10^{-2}$.
The overlapping region easily satisfies the $WZ/WW$ and $WH/ZH$ upper limits.
As we shall discuss the signal cross sections in the next section, 
the signal cross section for $\sigma(W') \times B(W' \to WZ)$ 
is of order $5-10$ fb while that for $\sigma(Z') \times B(Z' \to WW)
\alt 3$ fb.
We show the band of $5-10$ fb cross sections onto the Fig.~\ref{fig-para-w}.
The sweet spot is the strip obtained by overlapping 
the allowed region and the band of $5-10$ fb. 
While in Fig.~\ref{fig-para-z} we show the band of $2-5$ fb with a cyan curve
at 3 fb, because of the addition constraint 
$\sigma(Z') \times B(Z' \to W^+ W^- ) < 3 $ fb via semileptonic mode
\cite{cms-ww}.

\section{$W' \to WZ $ and $Z' \to W^+ W^-$ production }
The favorable region of parameter space in $g_R$ versus $\sin\phi_w$ 
is shown in Fig.~\ref{fig-para-w}. We can pick a point in the sweet spot
to account for the excess observed in the $WZ$ channel. From the 
ATLAS report, the number of excess events is about $8-9$ events 
around the 2 TeV peak. The selection efficiency for event topology
and boson-tagging requirements is about 13\% for a 2 TeV $W'$ boson
\cite{atlas}. With a luminosity of 20.3 fb$^{-1}$ it converts to 
$\sigma(W') \times B(W' \to WZ) \approx 6-7$ fb (here we take the
hadronic branching ratio of a $W$ boson or a $Z$ boson to be $0.7$).

\begin{figure}[th!]
\includegraphics[width=3in]{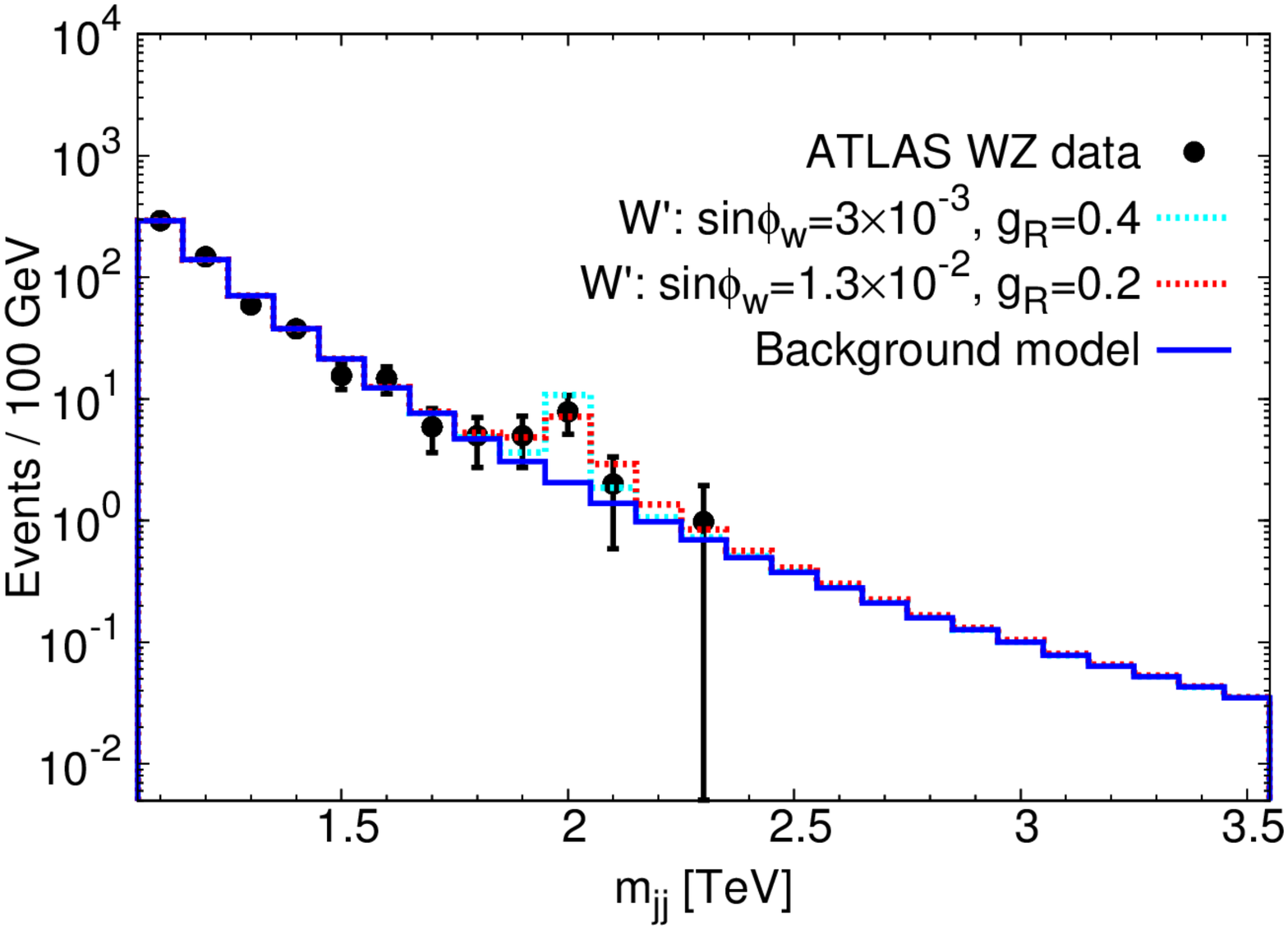}
\includegraphics[width=3in]{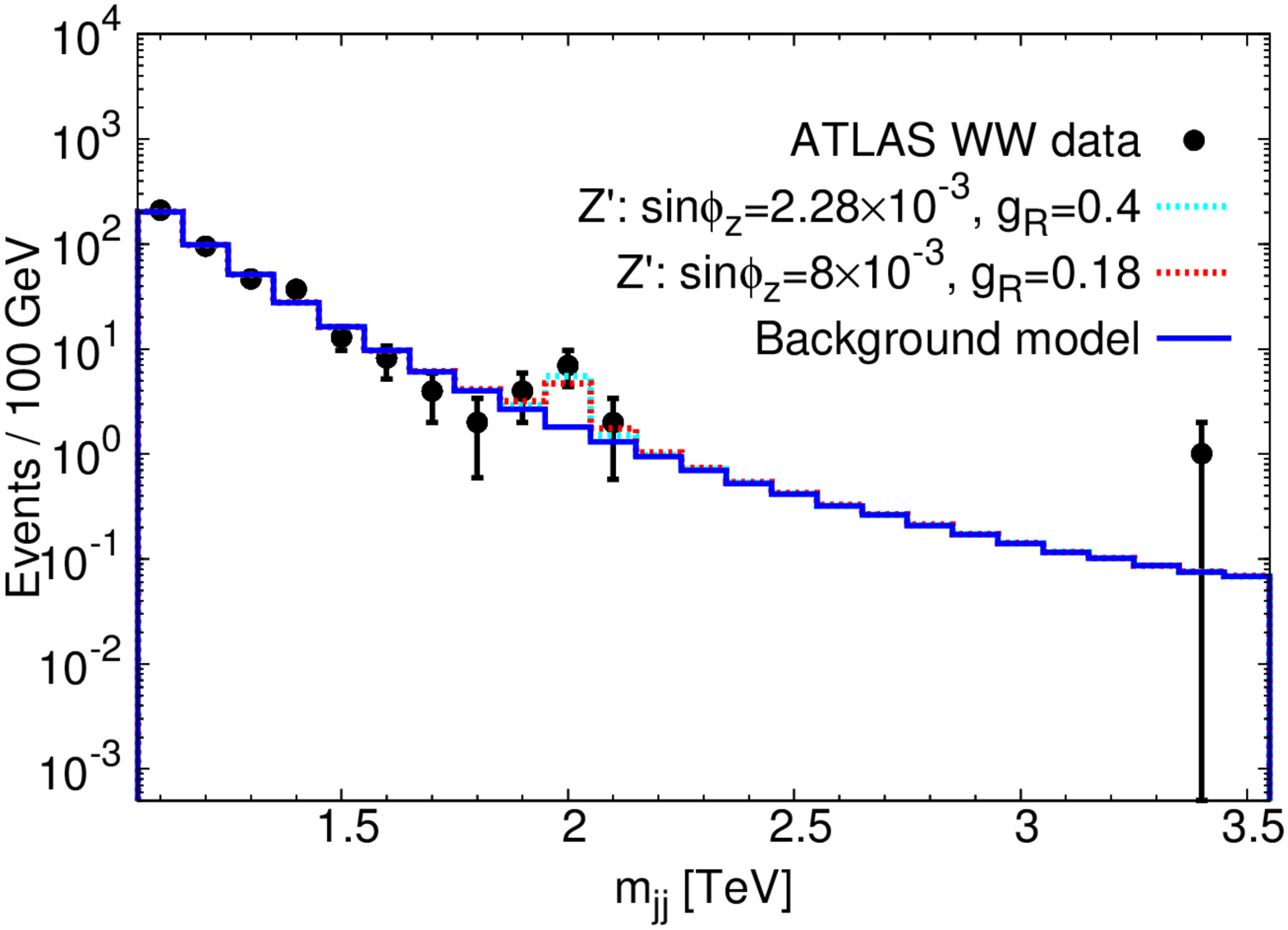}
\caption{\small \label{fig-w-wz}
Dijet invariant mass distribution for 
(left) $pp \to W'^\pm \to W^\pm Z$ and 
(right) $pp \to Z' \to  W^+ W^-$ 
with $M_{W'} = M_{Z'} = 2$ TeV. Here the finite width
effect is included.  A selection efficiency of $0.13$, hadronic branching
ratio of $0.7$ for each $W$ and $Z$ boson, and a luminosity of 20.3 fb$^{-1}$
are used. 
The dijet backgrounds are given in the ATLAS report \cite{atlas}.
}
\end{figure}

In Fig.~\ref{fig-para-w}, we show the band of the 
$\sigma(W')\times B(W'\to WZ) = 5- 10$ fb.
The sweet spot is the strip obtained by overlapping 
the allowed region and the band of ``$WZ:5-10$ fb''. 
Let us pick
a couple of representative points:
(i) $\sin\phi_w = 3 \times 10^{-3}$ and $g_R = 0.4$ (small mixing but
large $g_R$), and
(ii) $\sin\phi_w = 1.3\times 10^{-2}$ and $g_R = 0.2$ (large mixing but
small $g_R$). 
The mixing angle for the second point is at the upper limit allowed by
the EW constraint.
Then we calculate 
$\sigma(pp \to W'^\pm \to W^\pm Z)$ including the width effect,
and add to the dijet background shown in the ATLAS report \cite{atlas}.
We show the sum of the resonance peak and 
the dijet background in the left panel of Fig.~\ref{fig-w-wz}. 
Such a resonance contribution can explain the excess in the $WZ$ channel.
We can see that with small mixing but large $g_R$ (the cyan histograms)
the width of the 2 TeV resonance is narrower while that with large mixing
but small $g_R$ (the red histograms) is broader. Both choices can account 
for the data points within the uncertainties.

We repeat the exercise for the $Z'$.  The number of excess events
is about $7-8$ events around the 2 TeV peak. The selection efficiency
is about the same as the $W'$. It eventually converts to 
$\sigma(Z') \times B(Z' \to W^+ W^-) \approx 5-6$ fb. 
However, due to a recent search \cite{cms-ww} using semileptonic decay
mode, the 95\% CL limit on $\sigma(Z') \times B(Z' \to W^+ W^-) < 3$ fb.
Although there is a slight inconsistency, we pick a couple of 
representative points such that each gives a cross section about 3 fb:
(i) $\sin\phi_z = 2.28 \times 10^{-3}$ and $g_R = 0.4$, and
(ii) $\sin\phi_z = 8\times 10^{-3}$ and $g_R = 0.18$ from 
the sweet spot of Fig.~\ref{fig-para-z}.
Note that the mixing angle of the second point is at the 
upper limit allowed by the EW constraint.
We show the sum of the resonance peak and 
the dijet background in the right panel of Fig.~\ref{fig-w-wz}. 
Such a resonance contribution can roughly 
explain the excess in the $W^+ W^-$ channel within uncertainty.

\section{A unified $SU(2)_1 \times SU(2)_2 \times U(1)_X$ model}

Here we show that it is possible to have unified $W'$ and $Z'$ bosons in 
a model with an additional $SU(2)$ symmetry, and it will approach
the models of $W'$ and $Z'$ that we used in Sec. II, III and IV.
We follow closely the discussion in a couple of recent works addressing
the same anomaly \cite{221}.
We start with a popular scenario based on the symmetry
breaking pattern from the gauge group 
$SU(2)_1\times SU(2)_2\times U(1)_X$ 
(with gauge coupling $g,\, g'_2,\, g_X$ respectively), 
which is first broken into a lower symmetry 
$SU(2)_1\times U(1)_Y$ at the scale above TeV, and then broken again
at the electroweak scale \cite{vernon}. The intermediate symmetry is just 
the SM gauge group $SU(2)_L\times U(1)_Y$. 
The SM hypercharge convention 
is fixed by  $Q=T^{(1)}_3+ \frac{Y}{2} = 
T^{(1)}_3 + T^{(2)}_3 + \frac{Y_X}{2}$.

We choose the leptophobic version of the model such that the right-handed
$u_R$ and $d_R$ quarks are arranged in doublet of the $SU(2)_2$ while the
$\nu_R$ and $e_R$ as singlets of the $SU(2)_2$. The assignment of 
$T_3^{(2)}$ and $Y_X/2$ for the right-handed fermions are
\[
 \begin{array}{|c||c|c|c|c|} \hline
 f &   u_R &  d_R  &  \nu_R & e_R \\ \hline \hline
 T_3^{(2)} &  +\frac12  & -\frac12 &   0   &  0 \\ 
 \hline
 \frac{Y_X}{2}    &  + \frac{1}{6} & + \frac{1}{6} &  0 & -1 \\
\hline \end{array}    
\]

The first step of
symmetry breaking at TeV scale can occur via a Higgs doublet 
$\Phi \sim (1,2,1/2)$ under $SU(2)_1 \times SU(2)_2 \times U(1)_X$:
 \[
   \Phi = \left( \begin{array}{c}
                  \phi^+ \\
                  \phi^0  \end{array} \right),\qquad 
  \langle \Phi \rangle = \frac{1}{\sqrt{2}}\, \left( \begin{array}{c}
                   0  \\
                   u  \end{array} \right) \;.
\]
The gauge field $B'_\mu$ of the $U(1)_X$ and the $W^{'3}_\mu$ of the $SU(2)_2$
are rotated by angle $\phi$ into the $B_\mu$ of the $U(1)_Y$ and the
$Z'$ boson:
\[
 \left( \begin{array}{c} 
              B'_\mu \\
              W^{'3}_\mu \end{array}  
  \right ) = 
 \left( \begin{array}{cc} 
                  \cos\phi & - \sin\phi \\
                  \sin \phi &  \cos\phi  \end{array} \right ) \;
 \left( \begin{array}{c} 
              B_\mu \\
              Z^{'}_\mu \end{array}  \right )  \;.
 \]
while the second step is the usual breaking of the EW symmetry by
another Higgs doublet with a VEV $v$.
In order to obtain the coupling of the $B_\mu$ the same as the SM hypercharge
$g_1 Y/2 = (e/\cos\theta_w) Y/2 $ in the first step of 
symmetry breaking, we require
\begin{equation}
  g_X \cos\phi = g_1,\;\;\; 
  g'_2 \sin \phi = g_1,\;\;\; 
\tan\phi = \frac{g_X}{g'_2} ,\;\;\; \frac{Y_X}{2} + T_3^{(2)} = \frac{Y}{2}\;.
\end{equation}
The $W'$ and $Z'$ bosons obtain masses as 
\begin{equation}
  M^2_{W'} = \frac{ e^2 v^2}{4 \cos^2\theta_w \sin^2\phi} \, (x +1 ),\;\;\;
  M^2_{Z'} = \frac{ e^2 v^2}{4 \cos^2\theta_w \sin^2\phi \cos^2\phi} \, 
        (x + \cos^4 \phi ),
\end{equation}
where $x \equiv u^2 /v^2$ is very large. Therefore, in leading order
$M_{W'} \approx M_{Z'}$ if $\cos\phi \approx 1$. This is exactly the
limit that we want to pursue, and we shall show that the couplings of the
$W'$ and $Z'$ to fermions will approach the values that we used in 
the analysis. 

Note that $x \sim ( 2\,{\rm TeV} / 0.1 \,{\rm TeV} )^2 = 10^2 - 10^3$. 
The size of $\sin\phi$ cannot be much smaller than $0.3$ given $g'_2 \alt 1$.
In the limit of $x$ being large,
the left-handed and right-handed couplings of the $W'$ boson to SM
fermions become \cite{221}
\begin{equation}
\frac{ g_L^{W' f f'}} {g_R^{W' f f'} }
  \longrightarrow  \frac{1}{ x  } \;, \;\;\;\; {\rm with} \;\;
  g_R^{W' f f'} = \frac{g'_2 }{\sqrt{2}} \;,
\end{equation}
which is exactly the same as the $W'$ interaction in Eq.~(\ref{eq1}) with
$g'_2 = g_R$.
Similarly, in the limit of large $x$ and small $\sin\phi$, 
the left-handed and right-handed couplings of the $Z'$ boson 
to SM fermions become  \cite{221}
\begin{eqnarray}
 g_{f,l} &\longrightarrow &   \frac{g'_2}{\cos\phi} (T_3^{(1)} - Q )\, 
          \sin^2\phi \nonumber \\
 g_{f=\ell,r} &\longrightarrow & \frac{g'_2}{\cos\phi} ( - Q  \sin^2\phi )
    \nonumber \\
 g_{f=q,r} &\longrightarrow & \frac{g'_2}{\cos\phi} (T_3^{(2)} -
       Q  \sin^2\phi ) \nonumber
\end{eqnarray}
Note that the leptonic couplings $g_{f=\ell,l/r}$ are suppressed by 
$\sin^2\phi$ and also because its $T_3^{(2)} = 0$.
The left-handed couplings $g_{q,l}$ of quarks are also suppressed by
$\sin^2\phi$.  Therefore, only the right-handed couplings of quarks are
left unsuppressed, 
which is close to what we used in the analysis of $Z'$ 
with $g'_2 = g_R$ in previous sections. 
Therefore, in the limit of large $x$ we have more or less
achieved the leptophobic scenario with $W'$ and $Z'$ bosons
having a similar mass at 2 TeV and couplings to right-handed quarks only.

\section{Discussion}

We have considered a phenomenological $SU(2)_R$ model that contains
extra $W'$ and $Z'$ bosons, which mix with the SM $W$ and $Z$ bosons,
respectively. Thus, it can induce the decays of $W' \to WZ$ and $Z' \to 
W^+ W^-$ to explain the ATLAS anomaly in the diboson channels, while we 
interpreted the excess in $ZZ$ as a fluctuation or a substantial 
overlap with $WW$ and $WZ$. It is very difficult for a spin-1 boson to 
decay significantly into $ZZ$. 

We have applied the constraints of the total width of the $W'$ and $Z'$
bosons, dijet cross sections,  
$WZ$ and $WW$ cross sections, and 
$WH$ and $ZH$ cross sections for the $W'$ and $Z'$ bosons, respectively,
as well as qualitatively the EW precision constraints
on the parameter space of $g_R$ and the mixing angles $\phi_w$ and $\phi_z$.
We have found a sweet spot that satisfies all the constraints, and there exists
a viable region that can explain the excess in the $WZ$ and $W^+ W^-$ 
channels, respectively. The size of the mixing angle is 
$\phi_w, \phi_z \approx 3\times 10^{-3} - 10^{-2}$
 and the size of the coupling $g_R \approx 0.2 - 0.5$. 

We offer comments on our findings and other possibilities as follows.

\begin{enumerate}
\item The production of $WZ$ and $WW$ via $W'$ and $Z'$ bosons receives
a large enhancement due to the longitudinal polarization of the $W$ and 
$Z$ boson ($\epsilon_L^\mu(W/Z) \sim p^\mu /M_{W/Z}$).  If each boson-jet
system (which contains 2 closely separated jets) 
is boosted back to the rest frame of the $W/Z$ boson and the angle made
by the jet is measured, one may be able to tell the polarization of the
$W/Z$ boson.  

\item Another important channel to check is the semileptonic 
decays of the $W$ and $Z$ bosons, i.e.,
one boson decays leptonically while the other hadronically. Though the
event rates will be lowered, the $W$ or $Z$ peak can be easier 
distinguished.

\item As we have mentioned that it is very difficult to have a spin-1
  boson to decay into $ZZ$ at tree level. There are only
    two effective operators describing such vertex \cite{Keung:2008ve},
    one of which may be induced by anomaly associated with the extra
    $U(1)$ while the other must be CP violating.  The logical choice
  is spin-0 or spin-2.  However, the production of spin-0 boson, just
  like the SM Higgs boson, has to go through $gg$ fusion or $WW$
  fusion. The production cross sections are too small or the total
  decay width of the boson is too broad.  The spin-2 boson, e.g, the
  graviton Kaluza-Klein state of the Randall-Sundrum model, can decay
  into $WW$ and $ZZ$, but in the ATLAS report \cite{atlas} it was
  shown that the production rate of the spin-2 graviton is somewhat
  too small to explain the anomaly.

\item Another possibility is an extended Higgs sector. It is well-known
that in models with extra Higgs doublets the charged Higgs cannot
couple to $WZ$ at tree-level. It has to go beyond the doublet to e.g.
triplet models. One viable triplet model is the Georgi-Machacek 
model \cite{gm}
that contains neutral, singly-charged, and doubly-charged Higgs bosons
\cite{cw}.
The excess in $WW$ channel did not distinguish between $W^+ W^-$ 
and $W^\pm W^\pm$.
In particular, the doubly and singly charged 
$H_5^{++}$ and $H_5^+$ can be copiously produced via
vector-boson fusion for Higgs-boson mass at 2 TeV, but however the width
of the bosons are too broad to be consistent \cite{logan}.

\item Another  alternative is the strong dynamics \cite{tech2}, e.g., 
technicolor models. For example, a neutral $\rho_{TC}^0$  of 
2 TeV can decay into $W^+ W^-$ while a charged $\rho_{TC}^\pm$  of 
2 TeV can decay into $W^\pm Z$.

\end{enumerate}

\section*{\bf Acknowledgments}
We thank Abdesslam Arhrib for useful discussion.
W.-Y. K. thanks the National Center of Theoretical Sciences and 
Academia Sinica, Taiwan, R.O.C. for hospitality.
This research was supported in parts 
by the Ministry of Science and Technology (MoST) of Taiwan
under Grant Nos. 101-2112-M-001-005-MY3, 
102-2112-M-007-015-MY3 and by 
US DOE under Grant No. DE-FG-02-12ER41811.

%%%%%%%%%%%%%%%%%%%%%%%%%%%%%%%%%%%%%%%%%%%%%%%%%%%%%%%%%%

\end{document}